\documentstyle[aps,twocolumn]{revtex}

\begin {document}
\twocolumn[\hsize\textwidth\columnwidth\hsize\csname @twocolumnfalse\endcsname

\title{One-dimensional Kondo lattice model as a 
Tomonaga-Luttinger liquid}

\author{Naokazu Shibata} 
\address{
Institute for Solid State Physics, University of Tokyo,  
7-22-1 Roppongi, Minato-ku, Tokyo 106, Japan \\
}
\author{Alexei Tsvelik} 
\address{
Department of Physics, University of Oxford, 1 Keble Road,
Oxford, OX1 3NP, UK
}
\author{Kazuo Ueda}
\address{
Institute for Solid State Physics, University of Tokyo, 
7-22-1 Roppongi, Minato-ku, Tokyo 106, Japan
}

\date{3 Feb 1997}
\maketitle

\begin{abstract}
 Arguments are presented that in the one-dimensional Kondo lattice
model $f$-electron spins participate in filling of the Fermi sea. 
It is shown that in its paramagnetic phase this model belongs to 
the spin-1/2 Tomonaga-Luttinger liquid universality class. 
The ratio of the spin and charge velocities $v_{\sigma}/v_{\rho}$ 
and $K_{\rho}$ are  estimated to be of the order
of $(T_K/\epsilon_F)^{1/2}$. 
\end{abstract}
\vskip1pc]

\narrowtext
\section{Introduction}

 One can use a one dimensional Kondo Lattice (KL) model as
a toy model to study the long standing problem of whether  
localized electrons determine the volume of the Fermi surface. 
The numerical results \cite{shibata},\cite{caron} show that in 
the region of the phase diagram where KL  model  belongs to the 
universality class of spin-1/2 Tomonaga-Luttinger (TL) 
liquid\cite{Hal} the Friedel oscillations are characterized by the 
large Fermi vector. So it seems that the $f$-electrons do 
participate in the Fermi surface formation.  

 In this paper we undertake a further study of the TL phase of 
the KL model. Let the reader recall that the spin-1/2 TL liquid    
critical point is characterized by
two parameters: the ratio of spin to charge density wave velocities
$v_\sigma/v_\rho$ and the number  $K_\rho$ which parameterizes 
scaling dimensions
in the charge sector (the similar parameter in the spin sector is
fixed by the SU(2) symmetry, $K_\sigma = 1$). The dynamical 
spin and charge susceptibilities at low $(\omega, q)$ 
are given by 
\begin{eqnarray}
\chi_\sigma(\omega_n, q) =  \frac{2}{\pi}\frac{q^2}{q^2v_\sigma +
\omega_n^2/v_\sigma}\ , \label{resps}
\end{eqnarray}
\begin{eqnarray}
\chi_\rho(\omega_n, q) =  \frac{2K_{\rho}}{\pi}\frac{q^2}
{q^2v_\rho + \omega_n^2/v_\rho}\ . \label{resp} 
\end{eqnarray}
Thus if we shall manage to find these two parameters, the
characterization of the low energy sector of KL model is complete.

\section{Luttinger liquid parameters obtained by the $1/N$ 
expansion}

The Hamiltonian of 
the one-dimensional KL model is 
\begin{equation}
H  =  -t\sum_{i \sigma}
( a_{i \sigma}^\dagger a_{i+1 \sigma} + \mbox{H.c.})  
+J \sum_{i \mu} S^\mu_i s^\mu_i \label{Ham}
\end{equation}
where $a_{ i \sigma}^\dagger \  (a_{ i \sigma}) $ is the creation 
(annihilation) operator of a conduction electron  
at the $i$th site, and $ s^\mu_i=(1/2)\sum_{\sigma\sigma'}
a_{i\sigma}^{\dagger} \tau^\mu_{\sigma\sigma'}a_{i\sigma'} $,
with $\tau^\mu_{\sigma\sigma'}$ ($\mu = x,y,z$) being the 
Pauli matrices, are the spin density operators of the conduction 
electrons. The spin densities
are coupled to the localized spins $S^\mu_i$ through an 
antiferromagnetic exchange coupling $J$.

  In order to obtain analytical results we shall extend the symmetry
of the KL model to the SU(N) and  resort to the
$1/N$-expansion (see \cite{piers} and 
\cite{grilli}). The corresponding Lagrangian density is 
\begin{eqnarray}
{\cal L} & = & a^*_j[\partial_{\tau} + \hat\epsilon(x)]a_j +
f^*_j\partial_{\tau}f_j + \nonumber \\
& & \mbox{i}\lambda(f_j^*f_j - qN) - \frac{J}{N}(a^*_jf_j)(f^*_ka_k)\ .
\label{Lag}
\end{eqnarray}

 Here the dynamical field $\lambda(\tau, x)$  is introduced to 
enforce the local constraint of the fermion occupation number. 
The number $q$ remains finite when $N \rightarrow \infty$. 

 Next  we decouple the interaction by the
Hubbard-Stratonovich transformation:
\begin{equation}
- \frac{J}{N}(a^*_jf_j)(f^*_ka_k) \rightarrow N\frac{V^*V}{J} +
V(a^*_jf_j) + V^*(f^*_ja_j)\ .
\end{equation}
The resulting partition function  is gauge invariant:
\begin{eqnarray}
f_j(\tau,x) \rightarrow f_j(\tau,x)\mbox{e}^{\mbox{i}\phi(\tau,x)}
\ ,\nonumber\\
V(\tau,x) \rightarrow V(\tau,x)\mbox{e}^{- \mbox{i}\phi(\tau,x)}
\ ,\nonumber\\
\lambda(\tau,x) \rightarrow \lambda(\tau,x) - \partial_{\tau}
\phi(\tau,x)\ .
\end{eqnarray}
It is convenient to choose the  gauge where  the field $V$ is 
real. We choose the following parametrizations:
\begin{eqnarray}
V(\tau,x)  & = & V_0\sqrt{1 + [r(\tau,x)/V_0\sqrt N]} \ , 
\label{par} \\
\mbox{i}\lambda & = & T_K + \mbox{i} u /\sqrt{N}\ .
\end{eqnarray}
where $V_0$ is the saddle point value of $V$ which we shall 
determine later and $r$ is a new field chosen in such a way 
that its  measure of integration is trivial. 

 We shall expand the partition function around its saddle point:
\begin{equation}
V = V_0\ , \ \ \ \ \mbox{i}\lambda =  T_K\ .
\end{equation}
Expanding to the second order in $r$ we get
\begin{eqnarray}
{\cal L}& =& {\cal L}_0 + {\cal L}_{int}\nonumber\\
{\cal L}_0 &=& a^*_j[\partial_{\tau} + \hat\epsilon(x)]a_j +
f^*_j(\partial_{\tau} + T_K)f_j + V_0(a^*_jf_j + c.c.)\nonumber \\
& &  \\
{\cal L}_{int}& =& \frac{r^2}{4J} + \frac{r}{2\sqrt N}
(:a^*_jf_j + c.c.:)
+ \frac{\mbox{i} u}{\sqrt N}:f^*_jf_j: \label{fluct} \nonumber \\
\end{eqnarray}
where the dots mean that the average is substructed: $:A: \equiv A -
\langle A\rangle$ and the $r^2$ term comes from the expansion of 
the square root in the expression for $V$ (\ref{par}). 
The saddle point parameters $V_0$ and $T_K$ 
are determined self-consistently by
vanishing of the terms linear in $r$ and $u$:
\begin{eqnarray}
\frac{1}{N}\sum_{j = 1}^N\langle f^+_j(n)f_j(n)\rangle = q , 
\label{self1}\\ 
\frac{1}{N}\sum_{j = 1}^N[\langle a^+_j(n)f_j(n)\rangle + \langle 
f^+_j(n)a_j(n)\rangle] = -\frac{2V_0}{J}\ . \label{self}
\end{eqnarray}

 In the leading order in $1/N$ the spectrum is determined by the
saddle point. This gives us a great advantage because 
the saddle point describes the 
large Fermi surface.  The  single electron spectrum has the 
following well-familiar form:
\begin{eqnarray}
E_{\pm}(p) = [\epsilon(p) + T_K]/2 \pm 
\sqrt{[\epsilon(p) - T_K]^2/4 + V_0^2}\ .
\end{eqnarray}
Substituting the saddle point Green's functions into
Eqs. (\ref{self1}, \ref{self}) we get 
\begin{equation}
q = \rho(0)V_0^2/T_K, \: T_K = D\exp[- 1/\rho(0)J]
\end{equation}
where $\rho(0)$ is the bare density of states per one channel, 
$D$ is the bandwidth. The expression for the new Fermi vector is 
\begin{equation}
k_F = k_F^{(0)} + \pi q\ .
\end{equation}
It follows from this equation  that the charge susceptibility 
remains unaffected by the presence of the spins.

 Only the mode $E_-(p)$ crosses the chemical potential. 
Near the Fermi points the spectrum can be linearized:
\begin{equation}
E_-(p) \approx \pm v^*(p \mp p_F), \ \ \ \ v^* = v_F\rho(0) T_K q\ .
\end{equation}

 Now one can calculate the spin- and charge density response 
functions (\ref{resp}) directly. The calculation of the spin-spin 
correlation function is straightforward: in the leading order 
in 1/N the only contribution comes
from the polarization loop of two $G_f = \langle\langle f
f^+\rangle\rangle$-functions. The result  reproduces 
Eq.(\ref{resps}) with $v_{\sigma} = v^*$.

 Calculation of  the charge response function is more complicated. 
We chose the following approach: first we shall integrate out 
the high energy degrees of freedom in the
partition function and obtain the effective action for the low 
energy sector; then we shall bosonize this action and obtain the 
parameters of the TL liquid. To do the integration it is 
convenient to diagonilize the saddle point Hamiltonian and to 
express the fermionic operators in terms of the new annihilation 
operators $A_{\pm, j}(k)$ corresponding to excitations with the
dispersion $E_{\pm}(k)$ (since the transformation is
diagonal in the flavour indices, we shall omit them):
\begin{eqnarray}
a(k) = \sqrt\alpha_k A_+(k) + \sqrt\beta_k A_-(k)\ ,\nonumber\\
f(k) = -\sqrt\beta_k A_+(k) + \sqrt\alpha_k A_-(k)\ ,
\end{eqnarray}
where 
\begin{equation}
\alpha_k + \beta_k = 1,\ \ \ \ \beta_k = \frac{1}{2}\left\{1 -
\frac{[\epsilon(k) - T_K]}{\sqrt{[\epsilon(k) - T_K]^2 +
4V_0^2}}\right\}\ .
\end{equation}
Substituting these expressions into Eq.(\ref{fluct}) and omitting 
the terms containing only $A_+$ we get:
\begin{equation}
\int \mbox{d}x {\cal L}_{int} = L_1 + L_2 
\end{equation}
\begin{eqnarray}
L_1 & = & \sum_q\frac{r(-q)r(q)}{4J} \nonumber \\
& & + \frac{1}{2 \sqrt N}\sum_{k,q} 
\left[r(q)\left(\sqrt{\alpha_{k+q}\alpha_k}
-\sqrt{\beta_{k+q}\beta_k}\right) \right.\nonumber \\
& & \left.
- 2\mbox{i}u(q)\sqrt{\beta_{k+q}\alpha_k}\right] 
[A^*_{+,j}(k + q)A_{-,j}(k)+ c. c.]  \\
L_2 & = & \frac{1}{\sqrt N}\sum_{k,q}[A^*_{-,j}(k +
q)A_{-,j}(k)] \nonumber \\
& & [\sqrt{\alpha_F\beta_F}r(q) + \mbox{i}\alpha_F u(q)]
\end{eqnarray}
where $\alpha_F$, $\beta_F$  are taken at the Fermi surface: 
$\alpha_F \approx 1, ~\beta_F \sim T_K/D$. 

 Integrating over $A_+$ we get in  the leading order in $1/N$ the
following action for the fields $r$ and $u$: 
\begin{eqnarray}
S_{eff} &=& \frac{1}{2}\sum_{\omega,q}\Pi(\omega,q) \nonumber \\
& & [4r(-\omega,-q)r(\omega, q) + u(-\omega, -q)u(\omega, q)]\ .
\end{eqnarray}
 
 To get the effective action for the low-lying excitations we 
need to know the  function $\Pi(\omega, q)$ 
for the area around $q = 0$ and for $q = 2k_F$. The
result is $\Pi(0,0) = \rho(0)$. 

We bosonize the fermionic operators:
\begin{equation}
\sum_{k,j}A^*_{-,j}(k +
q)A_{-,j}(k) = \mbox{i} \sqrt{N/\pi}q\Phi_{\rho}(q)~~(|q| \ll k_F)
\end{equation}
where $\Phi_{\rho}$ is the charge field and integrate over $u$ and
$r$. Since $\beta_F$ is so small, 
 the largest contribution to the
effective action comes from the fluctuations of the $u$-field.
The
bosonized version of the effective action in the charge sector 
is given by 
\begin{equation}
S_{eff} = \int\mbox{d}\tau\mbox{d}x
[\frac{1}{2v^*}(\partial_{\tau}\Phi_{\rho})^2 +
\frac{1}{2\pi\rho(0)}(\partial_{x}\Phi_{\rho})^2]\ .
\end{equation}

 From this action one can derive the canonical expression for the
charge susceptibility and $K_{\rho}$. At least in the leading 
order in 1/N the result does not depend on N: 
\begin{equation}
v_{\rho} = \sqrt{v^*/\pi\rho(0)}\ ,\label{vc}
\end{equation}
and 
\begin{equation}
K_{\rho} = \sqrt{\pi\rho(0)v^*}\ .\label{krho}
\end{equation}

\section{Density matrix renormalization group study}

 In order to check the validity of the large $N$ results for the 
$N = 2$ case 
we numerically estimate the TL liquid parameters  making use of
the density matrix renormalization group (DMRG)\cite{DMRG}.
This method is the most suitable for studying  long range and
low-energy properties since it allows one to  study long chains 
iteratively enlarging system size
and to obtain the ground state wave function
with only small systematic errors, 
which can be estimated from the 
eigenvalues of the density matrix.
The obtained results are consistent with the 
above arguments and indicate 
$K_\rho \ll 1/2$ in the Kondo limit.

 Now we shall describe results of the numerical analysis of 
the model (\ref{Ham}). 
The paramagnetic metallic state of this  model, 
which is expected to be  a TL liquid, 
is realized only in the region of 
rather weak exchange coupling away from both the
half-filling ($n_c=1$) and the low carrier density limit
($n_c\rightarrow 0$).
The ground state is always insulating at half-filling
and ferromagnetic both in the strong coupling limit
($J\rightarrow \infty$) for general carrier densities 
($n_c\ne 1$) and in the low carrier density limit\cite{PD}.

We first calculate spin excitation gap $\Delta_s$ 
and difference of chemical potentials $\mu_+-\mu_-$
as a function of the system size $L$.
As expected, both $\Delta_s$ and 
$\mu_+-\mu_-$ ( Fig. 1 (a) and (b) for the case of
$n_c=2/3$ and $J=1.8t,2.0t$) 
vanish in the bulk limit ($L\rightarrow \infty$), 
which confirms that the paramagnetic phase
of the KLM is a TL liquid. 

The finite size corrections of 
$\mu_+-\mu_-$ and  $\Delta_s$ in Fig.~1
are related to the charge susceptibility and 
the spin velocity, respectively.
Since we have used open boundary conditions
$\Delta_s(L)=v_\sigma\Delta k(L)=v_\sigma\pi/L$,
and $\mu_+(L)-\mu_-(L)= \Delta n_c(L)/\chi_\rho=
2/(\chi_\rho L)$. 
The obtained values are shown in Table I. 
Once we have obtained $v_\sigma$ then we can 
calculate $\chi_\sigma$ through the relation
$K_\sigma=\pi v_\sigma \chi_\sigma /2 $, see Eq.~(\ref{resps}).
Because the SU(2) symmetry in the spin space
guarantees $K_\sigma =1$, rather large $\chi_\sigma$ 
is obtained as is shown in Table I. 
This large $\chi_\sigma$ is naturally
expected because there are macroscopic 
number of almost free spins in both weak and
strong coupling regions. 
The $f$ spins are almost but not exactly independent 
with each other: in the weak coupling region,
$L$ almost free $f$ spins, and in the strong coupling region,
$L(1-n_c)$ $f$-spins unpaired with conduction electrons.

Now we discuss the charge susceptibility.
In the strong coupling limit 
it tends to the value for the free spinless fermions;
$\chi_\rho^{-1}= \pi t\sin{(\pi-\pi n_c)}$.
On the other hand in the weak coupling limit
we expect a $J$ independent charge susceptibility 
as is predicted by the Gutzwiller type
variational calculations\cite{Rice};
$\chi_\rho^{-1}=\pi t\sin{(\pi n_c/2)}$.
The density $n_c=2/3$ is rather special in the sense 
that the values expected for the strong coupling limit and 
the weak coupling limit are the same.
Thus we expect $\chi_\rho$ depends only weakly on $J$.
In general, in the weak coupling limit  
we have an asymptotic form of the charge velocity which 
is proportional to $K_\rho$ as
\begin{equation}
v_\rho= 2 K_\rho t \sin{(\pi n_c/2)}
\end{equation}
from the relation 
$K_\rho=\pi v_\rho \chi_\rho /2$, Eq.~(\ref{resp}).
However, we have to be careful 
close to the half-filling where the charge 
susceptibility tends to diverge owing to the
charge gap at half-filling.

The estimation of the correlation exponent is 
one of the most difficult calculations even by
the DMRG method. In order to estimate $K_\rho$
we need to see long range behaviors of the
system with sufficient accuracy.
In the present study we use asymptotic form of 
the Friedel oscillations because they are 
numerically more reliable than long range off-diagonal
correlations.

The Friedel oscillations are density oscillations 
induced by a local perturbation. 
In a TL liquid, power low anomalies in 
correlation functions naturally reflect themselves 
in the Friedel oscillations;
the Friedel oscillations induced 
by an impurity potential are
\begin{eqnarray}
\delta \rho(x)& \sim & C_1\cos(2k_Fx)x^{(-1-K_\rho)/2}
 + C_2\cos(4k_Fx)x^{-2K_\rho} 
\nonumber \\
\label{CDO}
\end{eqnarray}
as a function of the distance $x$ from the impurity
\cite{wire1,wire2,wire3}, 
and analogously, spin density oscillations 
induced by a local magnetic field behave as 
\begin{eqnarray}
 \sigma(x) & \sim & D_1\cos(2k_Fx)x^{-K_\rho}.
\label{SDO}
\end{eqnarray}
Thus, we can determine $K_\rho$ from
the asymptotic form of the oscillations.

Fig.~2 shows induced charge- and spin-density Friedel oscillations 
of the KLM obtained by the DMRG for $J=2.5t$ at $n_c=6/7$.
The Fourier components of spin-density Friedel oscillations 
for $J=1.8t, 2.5t, 2.5t$ at $n_c=2/3,4/5,6/7$, respectively, 
are also shown in Fig.~3.
The charge density Friedel oscillations are
induced naturally by the open boundary conditions of the system
and the spin density oscillations are introduced  
by applying local magnetic fields at the both ends.
As is already shown for $n_c=4/5$ in the previous 
work\cite{shibata}, the period of the oscillations are explained by 
the assumption of the spin-1/2 TL liquid with the large Fermi 
surface, $k_F=\pi(1+n_c)/2$, which includes $f$ spin densities 
as well as the density of conduction electrons.

Now we calculate the correlation exponent, $K_\rho$.
In order to obtain $K_\rho$, we simply use
the slope of the envelope function of the charge density 
oscillations assuming that its decay is proportional to 
$x^{-2K_\rho}$, because dominant component of the 
oscillations is the $4k_F$ component even for the 
case of $J=1.5t$. 
In Fig.~4, the obtained $K_\rho$ for the exchange coupling 
from $J=4.0t$ to $1.5t$ at $n_c = 2/3$ are presented.
Since the $2k_F$ spin density oscillations
decay much slower than the charge density oscillations,
it is not possible to determine $K_\rho$ from 
the spin density oscillations in the present system size.
However, the slower decay of the spin density oscillations
is consistent with the TL liquid prediction, Eq.~(\ref{SDO}), 
which gives smaller exponent, $x^{-K_\rho}$.

As is clearly seen in Fig.~4, $K_\rho$ is always smaller than
$1/2$ and monotonically decreases with decreasing $J$.
In the strong coupling limit, the conduction electrons and
the localized $f$ spins form local singlets leading to a
complete spin-charge separation.
Since the charge part is described by the free spinless fermions,
$K_\rho=1/2$ is obtained in the strong coupling limit
as in the case of the infinite-$U$ Hubbard model.
With decreasing $J$ from infinity the repulsive interaction 
between the neghboring spinless fermions is introduced 
in the leading order of $t/J$. 
Thus the situation is similar to the 
large-$U$ Hubbard model with nearest neighbor repulsions
whose $K_\rho$ is smaller than $1/2$\cite{Hub,NUH}.

In Fig.~4 we find a small discontinuity at $J=2.4t$. 
This is due to the phase transition from the 
ferromagnetic state to the paramagnetic one. 
Since this transition is of the first 
order accompanied by a jump in the total spin quantum number, 
from $S=L(1-n_c)/2$ to $0$, or $1/2$ with $L$ being the number 
of the sites, it is natural that the $K_\rho$ also shows a jump
at the critical value $J_c$. 
In order to confirm the discontinuity we have calculated the 
$K_\rho$ in both ferromagnetic and the paramagnetic states 
at $J=2.4t$ which is near but smaller than the critical point.
The $K_\rho$ in the ferromagnetic state is calculated 
by setting the total $S_z$ being $L(1-n_c)/2$ which is 
the total spin in the ferromagnetic state. 

In contrast to the slow decrease of $K_\rho$ above the 
critical $J_c$, a rather sharp decrease is observed 
below $J_c$, and the $K_\rho$ becomes smaller than $1/3$ 
which means that the long range behavior of the 
charge-charge correlation is governed by the $4k_F$ oscillations. 
The dominance of the $4k_F$ oscillations is 
a characteristic feature of this new class of spin-1/2 TL liquid. 

With further decreasing $J$, the $K_\rho$ seems to cross 
the value $3-2\sqrt{2} \sim 0.17$. 
Since the exponent of the power low anomaly in the momentum 
distribution function, $\alpha$, is given by
$\alpha=(K_\rho+1/K_\rho-2)/4$,
the power low anomaly is removed below this point 
and we cannot see clear Fermi surface any more. 
It is very difficult to observe clear Friedel oscillations 
for smaller $K_\rho$ than $0.17$.

\section{Conclusions}

 In conclusion we have established that in the area of phase 
diagram where 
the one dimensional Kondo lattice is paramagnetic, it belongs to the
universality class of spin-1/2 TL liquids. The $f$-electrons
do take part in formation of the Fermi surface. According to the
Luttinger theorem the volume of the Fermi sea is determined by those
branches of the spectrum which cross the chemical
potential. Despite of the fact that most of the spectral weight 
of the $f$-electrons is concentrated far from the chemical 
potential, they do
have access to it via the Kondo resonance. We bring attention of the
reader to the fact that the Luttinger theorem does not require
existence of a pole in the single electron Green's function and
therefore can be applied outside of the Fermi liquid domain. In
particular, the system under
consideration belongs to the spin-1/2 TL liquid universality class. 
It is a rather peculiar member of this class since $K_{\rho}$ is 
small. How small is not entirely clear; the  analytical calculations
give $K_{\rho} = v_{\sigma}/v_{\rho}$ 
(see Eqs.(\ref{vc}, \ref{krho}))
which is one order of magnitude smaller than 
the values obtained numerically (see Table I).  
This may be due to the
inaccuracy of the $1/N$-approximation; it is more  likely  however,
that the maximal system size available for the numerical 
calculations
is not big enough to penetrate to the asymptotic region. Thus the
numerical values of $K_{\rho}$ given in the Table 
should be considered as upper
limits.  This may appear unusual
to those who consider the Hubbard model as a typical example of TL
liquid. The smallness of  $K_{\rho}$ clearly originates from the
nonlocality of the effective interactions in space and time. 
In this sense the KL model
is similar to Charge Density Wave systems 
where the interactions are also retarded being 
carried by low energy optical phonons. In these systems 
$K_{\rho} \ll 1$ \cite{br},\cite{fl}. 

\acknowledgments
A.~M.~T. and N.~S. acknowledge a support of Japan 
Society for Promotion of Science. 
A.~M.~T. also acknowledges a kind hospitality of  
Institute for Solid State Physics, University of Tokyo 
where significant part of this work was accomplished. 
He is also grateful to Gilbert Lonzarich, David Khmelnitskii 
and Andrew Schofield for interesting discussions and interest 
to the work. N.~S.\ also thanks Manfred Sigrist for valuable 
discussions.

\begin{table}
\caption{Luttinger liquid parameters of the one dimensional Kondo
lattice model. The carrier density $n_c$ is $2/3$.  The energy 
unit is $t$. The errors are estimated from the ambiguity of the 
power low decay of the charge density Friedel oscillations. 
}
\begin{tabular}{l|ccccc} 
   & $K_\rho$ & $v_\sigma$ & $\chi_\sigma$ & 
                         $v_\rho$ & $\chi_\rho$ \\ \hline
$J=0   $ & 1    & -     & -    & 1.73 & 0.37 \\
$J=1.5t$ & 0.19 $\pm$ 0.03 &       &    & 0.30 $\pm$ 0.06 & 0.42 \\
$J=1.8t$ & 0.24 $\pm$ 0.02 & 0.014 & 46 & 0.41 $\pm$ 0.06 & 0.38 \\
$J=2.0t$ & 0.27 $\pm$ 0.02 & 0.011 & 56 & 0.48 $\pm$ 0.06 & 0.36 \\ 
\end{tabular}
\label{table-1}
\end{table}

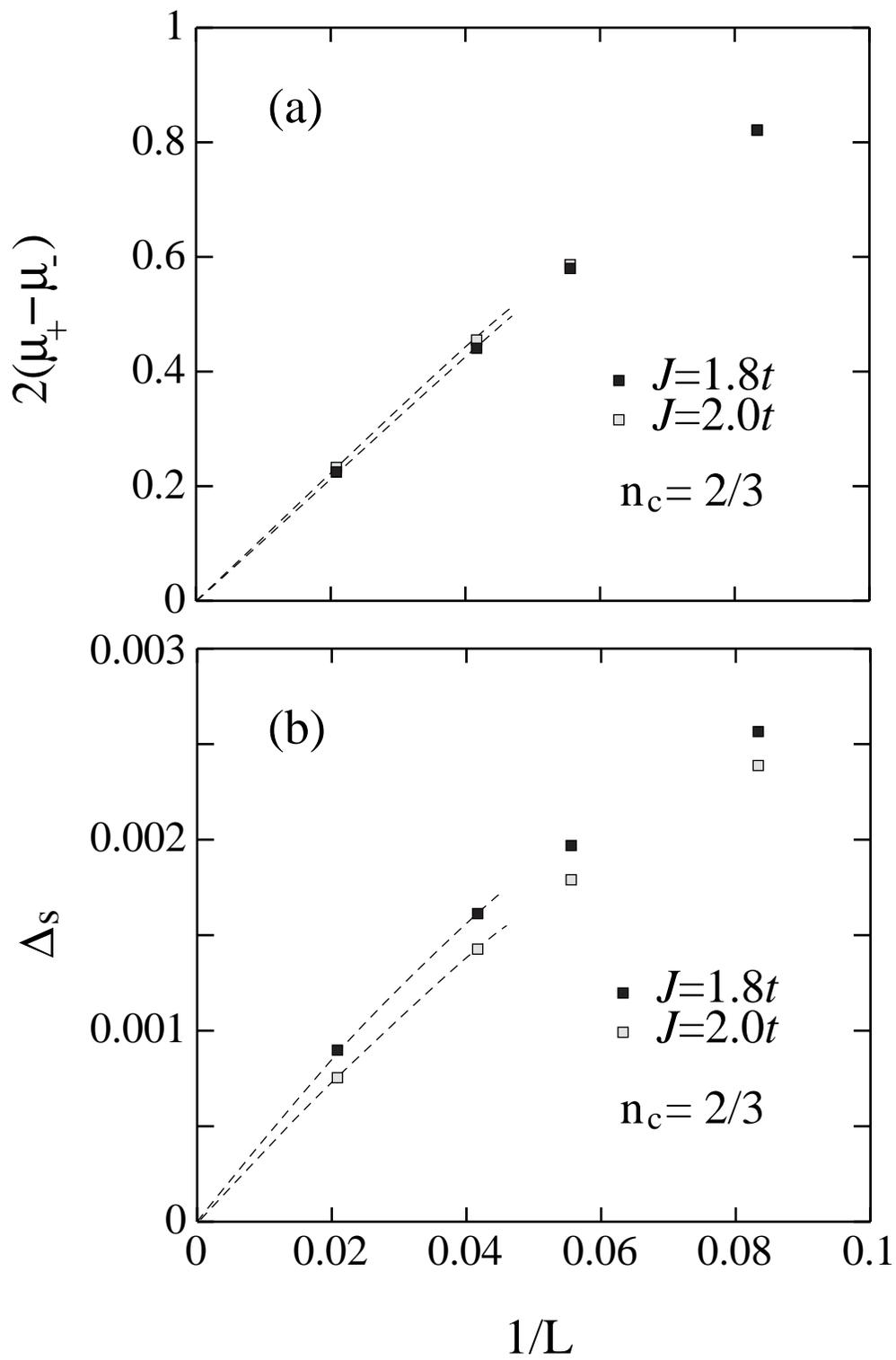
\begin{figure}
\caption{(a) Size dependence of the difference of the chemical 
potentials, $\mu_+-\mu_-$, in the one dimensional Kondo lattice 
model. 
$2\mu_+(L)=E_g(n_c=n_c^0 +2/L,L)- E_g(n_c=n_c^0,L)$. $2\mu_-(L)= 
E_g(n_c=n_c^0,L)- E_g(n_c=n_c^0 -2/L,L)$. $E_g(n_c,L)$ is the 
ground state energy at the carrier density $n_c$ in the system of 
length $L$. $n_c^0=2/3$.
(b) Size dependence of the spin gap 
$\Delta_s(L)=E_g(S_z^{tot}=1,L)- E_g(S_z^{tot}=0,L)$. 
$E_g(S_z^{tot},L)$
is the lowest energy in the Hilbert space of total spin $S_z^{tot}$.
$n_c=2/3$. The energy unit is $t$. 
Typical truncation errors in the DMRG calculations are $10^{-4}$.
}
\label{ChageGap}
\end{figure}

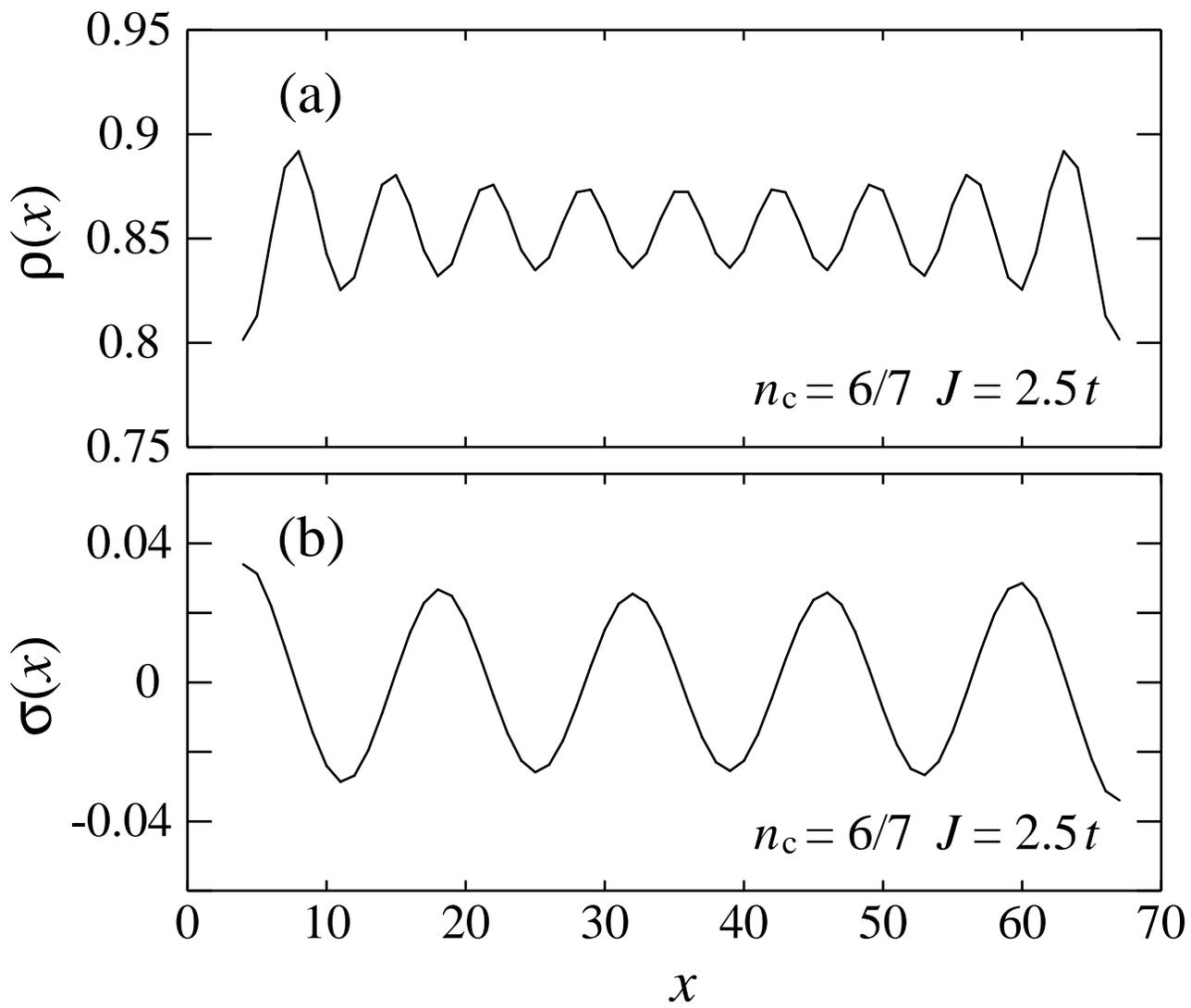
\begin{figure}
\caption{(a) Charge density Friedel oscillations induced by the open
boundary conditions. The system size is 70 sites. 
(b) Spin density Friedel oscillations induced by applying local 
magnetic fields at the both ends. The strength of the local magnetic 
fields is $0.2t$. Typical truncation errors in the DMRG calculations 
are $1 \times 10^{-6}$ for $J=2.5t$.
}
\label{CSFO}
\end{figure}

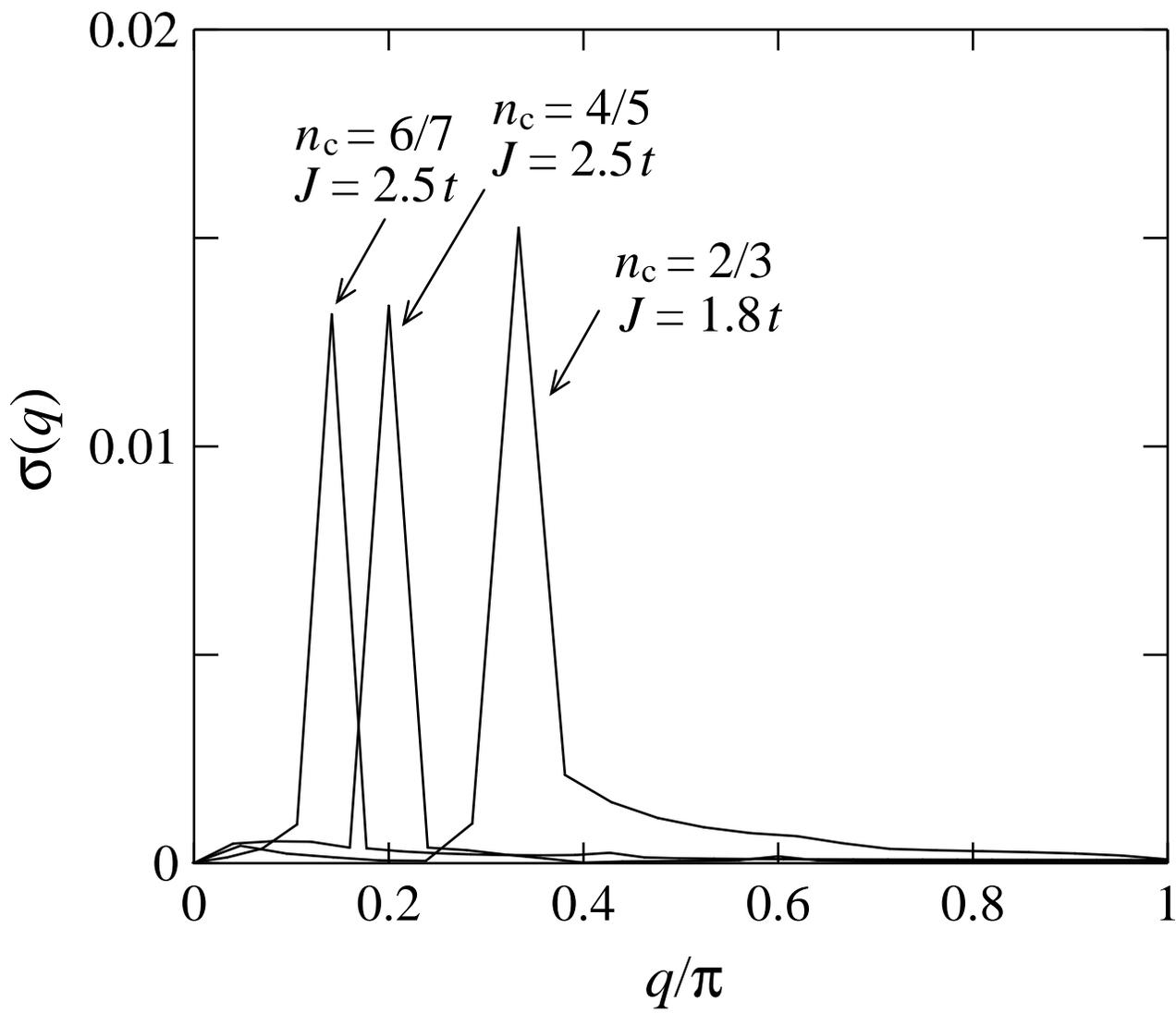
\begin{figure}
\caption{
Fourier components of the spin density Friedel oscillations. 
}
\label{FSFO)}
\end{figure}

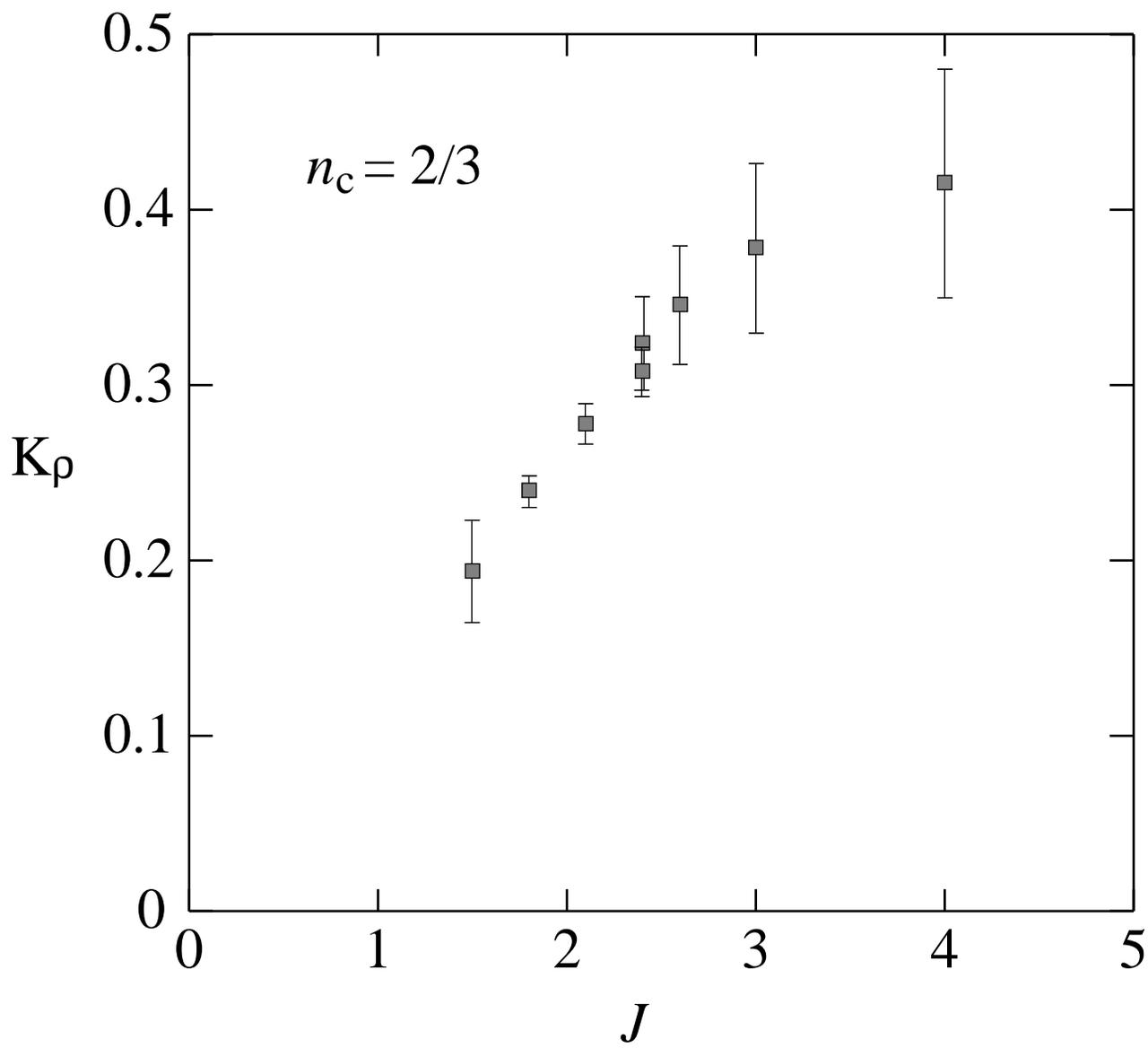
\begin{figure}
\caption{
Correlation exponent $K_\rho$ estimated from the decay rate
of the charge density Friedel oscillations.
The errorbars are determined from the ambiguity of the 
power low fitting. $n_c=2/3$. $J$ is in units of $t$.
}
\label{Kc}
\end{figure}

\end{document}